\documentclass[graybox]{svmult}

\usepackage{mathptmx}       % selects Times Roman as basic font
\usepackage{helvet}         % selects Helvetica as sans-serif font
\usepackage{courier}        % selects Courier as typewriter font
\usepackage{type1cm}        % activate if the above 3 fonts are
\usepackage{makeidx}         % allows index generation
\usepackage{graphicx}        % standard LaTeX graphics tool
\usepackage{multicol}        % used for the two-column index
\usepackage[bottom]{footmisc}% places footnotes at page bottom

\makeindex             % used for the subject index
\begin{document}

\title*{Infrared Emission from Supernova Remnants: Formation and Destruction of Dust}
\author{Brian J. Williams and Tea Temim}
\institute{Brian J. Williams \at CRESST/USRA and X-ray Astrophysics Laboratory, NASA/GSFC, Code 662, Greenbelt, MD, USA, \email{brian.j.williams@nasa.gov}
\and Tea Temim \at CRESST/UMCP and Observational Cosmology Laboratory, NASA/GSFC, Code 665, Greenbelt, MD, USA \email{tea.temim@nasa.gov}}
\maketitle

\abstract{We review the observations of dust emission in supernova
  remnants (SNRs) and supernovae (SNe). Theoretical calculations
  suggest that SNe, particularly core-collapse, should make
  significant quantities of dust, perhaps as much as a solar
  mass. Observations of extragalactic SNe have yet to find anywhere
  near this amount, but this may be the result of observational
  limitations. SN 1987A, in the process of transitioning from a SN to
  an SNR, does show signs of a significant amount of dust forming in
  its ejecta, but whether this dust will survive the passage of the
  reverse shock to be injected into the ISM is unknown. IR
  observations of SNRs have not turned up significant quantities of
  dust, and the dust that is observed is generally swept-up by the
  forward shock, rather than created in the ejecta. Because the shock
  waves also destroy dust in the ISM, we explore the question of
  whether SNe might be net destroyers, rather than net creators of
  dust in the universe.}

\section{Introduction}
\label{intro}

Supernova remnants (SNRs) provide a laboratory for the study of
various aspects of the evolution of the interstellar medium (ISM)
across the electromagnetic spectrum. SNRs are often bright in the
infrared (IR), where both continuum emission from warm dust and line
emission from molecules and low ionization atomic states of various
elements are observed. The ejecta from SNRs enrich the ISM of the
Galaxy with metal-rich products such as silicon, iron, oxygen, and
magnesium. These are the same elements that are predominantly found in
dust grains, leading to the conclusion that grains might be formed
within the expanding, cooling ejecta.

Dust plays an important role in all stages of galaxy evolution, as the
life cycle of dust grains and the amount and relative abundances
present in the ISM are determined by the balance between grain
formation, modification, and destruction. Dust heated in the powerful
shock waves of SNRs radiates strongly at IR wavelengths, so
IR observations provide a means to study both the properties of ISM
dust and the formation and destruction mechanisms in ways that are not
otherwise feasible. The conclusion that SNe are responsible for
creating most of the dust in the universe is not a closed case. During
the remnant phase, the shock wave races out into the ISM, heating and
compressing the interstellar or circumstellar gas. This gas, heated to
an X-ray emitting plasma of several million degrees, collides with
dust grains embedded within the medium, heating them to temperatures
of up to a few hundred degrees, but also modifying and destroying them
in the process. Even the newly formed grains within the ejecta are
subject to destruction by the reverse shock wave, and it may be that
SNe, on the whole, destroy more dust than they create.

In this work, we review some of the basic theoretical and
observational work behind IR emission in SNe and SNRs. We discuss the
heating and cooling mechanisms for gas and dust in remnants, and
discuss dust formation in SN ejecta. In general, the amount of dust
detected during the SN phase is quite small, only a few thousandths of
a solar mass. Most SNe fade from view in the IR within a few months,
though some, like Type IIn SNe, remain bright for years
post-explosion, presumably due to a significant amount of dust created
in the pre-SN outflows. Young SNRs (where young is loosely defined as
a few hundred to a few thousand years old) are typically bright mid-IR
sources, and {\it Spitzer} has returned many spectacular images and
spectra of these remnants. The shape of the dust spectrum can serve as
a diagnostic of both the grain properties, and the hot, X-ray emitting
plasma. We also discuss the destruction of dust by the forward and
reverse shock waves.

We summarize recent progress made in these areas, which has grown by
leaps and bounds over the past decade with the remarkable success of
the {\it Spitzer} and {\it Herschel} missions. The focus is on
Galactic and Magellanic Clouds (MCs) observations of SNRs, because the
resolved nature of remnants at these distances allows for detailed
study of the interaction of the shock wave with dust in the ambient
Galactic medium. While core-collapse SNe do show some evidence for
dust formation in their ejecta, Type Ia SNe do not. We also discuss
extragalactic observations of SNe, and finally, address the question
of whether SNe, as a whole, are net producers or destroyers of dust in
the universe.

\section{IR Emission from SNRs}

SNRs have several emission mechanisms, both thermal and nonthermal, in
the IR. Thermal emission can manifest as either a continuum from warm
dust grains or as line emission from atomic or molecular gas. Rarely,
nonthermal emission is present in the form of synchrotron emission
from relativistic electrons accelerated by shock waves. These are not
mutually exclusive, and some remnants have emission from all
three. The dominant emission mechanism depends on several factors, but
the primary determinant is the evolutionary stage of the remnant,
which can be parameterized as the product of the remnant's age and the
density of the material into which the forward shock wave is
expanding.

In general, remnants at an advanced evolutionary stage (either because
they are older or have encountered a denser medium) are dominated by
line emission from atomic and molecular gas. The gas temperatures in
these have cooled sufficiently to allow recombination to take place,
and the atomic fine structure lines from common elements such as Fe,
O, Si, and Ne dominate the spectrum. In shocked molecular gas, H$_{2}$
is the dominant coolant. We refer the reader to Reach et al. (2006)
for a more thorough review of the IR line emission in advanced
SNRs. These remnants are also bright in optical wavebands, showing up
as radiative shocks.

In young remnants, on the other hand, shock waves are still mostly
non-radiative, meaning that the plasma temperature in the post-shock
environment is still too high (the thermal emission from this gas is
in the X-ray band) for recombination to occur. As a result, these
remnants emit very little in optical wavebands, but can emit
significant IR emission as thermal continuum from dust grains, warmed
by collisions with energetic particles in the post-shock gas (we
detail the heating mechanism for this in Section~\ref{heating}). In
some young remnants, this thermal dust continuum is the {\it only}
source of emission in the IR. For examples of this, see Williams et
al. (2011).

\begin{figure}[h]
\includegraphics[scale=0.5]{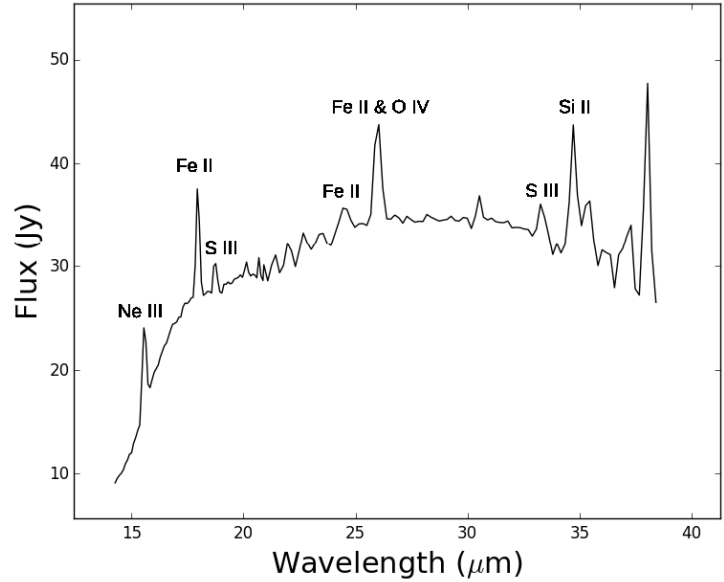}
\caption{{\it Spitzer} mid-IR spectrum of Kepler's SNR. The continuum
  results from warm dust at $\sim 100$ K. Several emission lines from
  low-ionization state ions are visible.}
\label{keplerspec}
\end{figure}

Most remnants show a mix of continuum and line emission sources,
though the exact physical locations of their origins may be
distinct. As an example of this, we show the integrated spectrum from
Kepler's SNR in Figure~\ref{keplerspec}. The spectrum contains strong
continuum emission and several bright emission lines. Williams et
al. (2012) interpreted the continuum as arising from the fast
non-radiative shocks, while the line emission comes from the slower,
radiative shocks, seen in the optical. Finally, some remnants, like
Cas A, are bright enough that a faint contribution from synchrotron
emission is detectable.

\subsection{Heating Mechanisms for Dust in SNe and SNRs}
\label{heating}

Dust grains in the ISM are microscopic, ranging in size from a few
nanometers to several microns, and are generally composed of
refractory elements, such as carbon, magnesium, silicon, and iron. On
average, about 0.1-1\% of the mass of the ISM is contained in dust
grains, with the remainder being in the gaseous phase (the bulk of
that being hydrogen and helium). ISM grains have a steep distribution
in size, roughly following a power-law given by $N(a) \propto
a^{-3.5}$, where $a$ is the grain radius.

In general, dust grains are heated in one of two ways: by absorption
of radiation or by direct collision with energetic particles. These
two mechanisms are not mutually exclusive. Most dust heating in the
universe is done via radiation. In the ambient ISM, grains are heated
by the interstellar radiation field, primarily ultraviolet starlight,
to temperatures of 10-20 K. Dust much closer to bright sources, such
as stars, can be heated to significantly higher temperatures, and is
often seen in mid-IR observations. Regardless of the heating
mechanism, dust grains will cool via radiation in the IR as a modified
blackbody, where the exact shape of the spectrum depends on the
optical properties of the grains themselves.

In SNe, there are several radiative mechanisms which can heat
dust. The simplest mechanism is the extreme luminosity from the SN
itself, powered by either radioactive decay or circumstellar
interaction of the shock wave and the dense material it is plowing
into. Grain temperatures observed in SNe can exceed 500 K, so
observations of dust in SNe must be done in the near-IR.

\begin{figure}[h]
\includegraphics[scale=0.32]{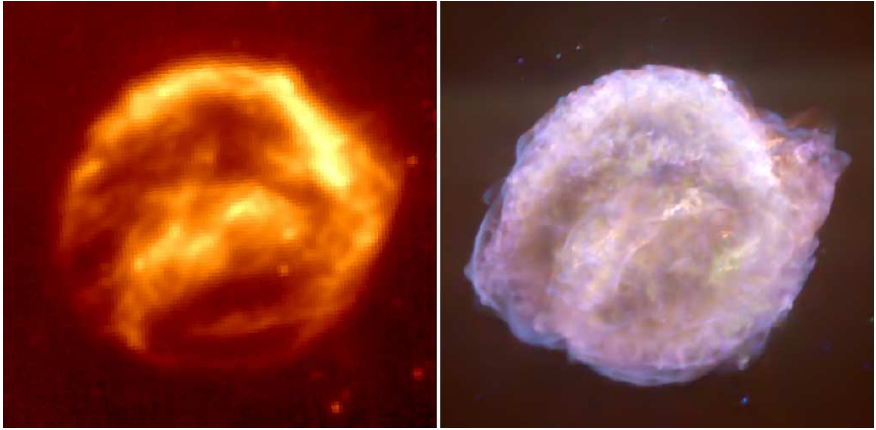}
\caption{{\it Left:} {\it Spitzer} 24 $\mu$m image of Kepler's SNR
  (PI: W.P. Blair). {\it Right:} {\it Chandra} 3-color mosaic (PI:
  S.P. Reynolds), with 0.3-0.72 keV emission in red, 0.72-1.7 keV
  emission in green, and 1.7-7.0 keV emission in blue. The
  morphological connections between the two wavebands are obvious; see
  the text for details. The image is approximately 5$'$ on a side.}
\label{kepler}
\end{figure}

In young remnants, the bulk of the radiation emitted by the shocked
plasma is at X-ray energies. But since dust grains are not efficient
absorbers of X-rays, the heating of grains in these remnants is
dominated by collisions with hot electrons and ions (Draine \&
Salpeter 1979), the same electrons and ions responsible for the X-ray
emission observed behind the forward shock in SNRs. This physical
connection between the X-ray and IR emission mechanisms accounts for
the similar morphologies of young remnants seen in these two wavebands
(Williams et al. 2006). See Figure~\ref{kepler} for an example using
the remnant of Kepler's SN of 1604 A.D.

In the late SNR phase, slow, radiative shocks ($< 150$ km s$^{-1}$)
can emit significant amounts of optical/UV radiation due to the
rapidly cooling gas in the post-shock environment. This radiation
generally becomes the dominant heating source in older remnants, since
young remnants are usually dominated by non-radiative shocks. The dust
in older remnants tends to be cooler, generally in the 20-50 K range,
emitting mostly in the far-IR. Virtually all of the emission from
older remnants is dominated by swept-up ISM dust.

The heating rate for a grain immersed in a hot plasma is given by

\begin{equation}
H=\left(\frac{32}{\pi m}\right)^{1/2}\pi a^{2} n (kT)^{3/2} h(a,T),
\end{equation}

\noindent
where $m$ is the mass of the impinging particle (proton, electron,
etc.), $a$ is the radius of the grain, $n$ is the density of the gas,
$kT$ is the temperature of the gas, and $h(a,T)$ is a function that
describes the efficiency of the energy deposition rate of a particle
at a given $T$ for a grain with radius $a$. It can be immediately seen
from this equation that at a given temperature, electrons will
dominate the heating over protons, since their mass is much smaller
and they move much faster. Young remnants, such as Kepler's SNR, have
grain temperatures in the 50-200 K range, and emit most of their light
between 20-50 $\mu$m. If the temperature of the plasma is known
reasonably well from X-ray observations, the density of the post-shock
gas can be used as a free parameter to match the IR spectral energy
distribution of the dust. For instance, in Kepler's SNR, a combined
IR/X-ray analysis showed that the density of the medium behind the
forward shock is $\sim 50$ cm$^{-3}$, orders of magnitude above what
one would expect for the ``typical'' ISM at Kepler's location in the
Galaxy (Williams et al. 2012), further solidifying the case that
Kepler is a Type Ia remnant with circumstellar interaction.

The radiative heating of grains by a nearby bright optical/UV source
is given by

\begin{equation}
H=\frac{\pi a^{2} \int f(\nu, a) L_{\nu}(\nu) Q_{abs}(\nu, a) d\nu}{4\pi r^{2}},
\end{equation}

where $f(\nu, a)$ is the fraction of energy deposited into the grain,
$L_{\nu}$($\nu$) is the luminosity of the source, $Q_{abs}$ is the
absorption coefficient for a given grain composition (a quantity which
depends on the optical properties of the grains), and $r$ is the
distance between the source and the dust grains. Regardless of the
heating source, grains cool by emitting radiation with a rate given by

\begin{equation}
L_{gr}(a) = 4\pi a^{2} \int \pi B_{\nu}(\nu, T) Q_{abs}(\nu, a) d\nu,
\end{equation}

where $B_{\nu}(\nu, T)$ is the Planck blackbody function with $T$ as
the grain temperature.

\section{Dust Formation in SNe}
\label{formation}

Observations of high-redshift quasars at far-IR and sub-millimeter
wavelengths revealed the presence of high masses of dust of $> 10^8
M_{\odot}$, and implied that this dust had to form on relatively short
timescales of a few hundred million years (see review by Gall et
al. 2011). Since core-collapse SNe occur on short enough timescales,
they were postulated to be the likely sources of dust in these young
galaxies. Other possible mechanisms include the efficient growth of
dust in the ISM (e.g. Draine 2009), and dust formation in outflowing
quasar winds (Elvis et al. 2002). However, studies of extinction
curves of high-redshift galaxies find that the flat shape of the
curves in the optical and UV can be explained by dust grains processed
by the SN reverse shock, providing further indication that SNe are the
dominant contributors.

Dwek et al. (2009) used dust evolution models to fit the spectral
energy distributions (SEDs) of high-redshift quasars, and determined
that an average SN would need to produce 0.1-1.0 $\rm M_{\odot}$ of
dust to explain the observations, assuming that none of the grains are
destroyed in the ISM. If dust destruction in the ISM is taken into
account (see Section~\ref{destruction}), the required mass would be
even larger. For this reason, characterizing the masses and properties
of dust grains produced in SNe, as well as the efficiency of their
eventual destruction in shocks, is critically important to understand
the origin of dust in the early and present day universe.

While most of the non-stellar, baryonic mass in the ISM is hydrogen
and helium, dust is comprised mostly of heavier metals, such as
silicon, iron, oxygen, and carbon. These elements are found, in high
abundances, in SN ejecta. Theoretical models for dust formation in SNe
predict that significant masses of dust can form in these ejecta over
the timescale of a few years post-explosion, but that the mass,
composition, size, and ultimate survival of dust grains strongly
depend on the type of the explosion and the metallicity (Kozasa et
al. 2009). Models for dust formation in core-collapse SNe based on
classical nucleation theory and the chemical kinematic approach for
the formation of molecular precursors predict that 0.03-0.7
$M_{\odot}$ of dust can form in the ejecta, with some of the most
abundant species being MgSiO$_{3}$, SiO$_{2}$, Mg$_{2}$SiO$_{4}$, and
C. However, only a fraction of this dust is expected to survive the SN
reverse shock (see Section~\ref{reverseshock}). The survival of dust
grains is primarily determined by their size. Kozasa et al. (2009)
found that Type IIb SNe with less massive envelopes, and therefore
lower densities and higher velocities in the ejecta, tend to form
smaller grains ($<$ 0.006 microns), while Type IIP SNe with massive
envelopes form relatively large grain ($>$ 0.03 microns). Sarangi \&
Cherchneff (2015) explored dust formation in Type IIP SNe by
accounting for the gas-phase chemistry, nucleation, and coagulation of
grains, and confirmed the result that these explosions tend to form
larger grains, especially if the ejecta are clumpy. They also found
that clumpy ejecta lead to somewhat higher dust masses, and a higher
fraction of metallic grains.

For SNe with a significant amount of pre-SN mass loss, it may be
possible for dust to form in a cool, dense shell that forms in between
the forward-shocked CSM and the reverse-shocked ejecta. Smith et
al. (2007) report observations of the Type Ib SN 2006jc, which shows a
rising near-IR continuum that is well-fit by hot graphite dust grains,
in combination with progressively more asymmetrically blueshifted He I
lines resulting from increasing dust obscuration of receeding
material. Interestingly, these effects were seen at very early times,
between 50 and 75 days, as opposed to the several hundreds of days
necessary for dust to form in the ejecta.

Nozawa et al. (2011) modeled the formation of dust in Type Ia SNe and
found that the total dust mass yield varies from 3 $\times 10^{-4}$ to
$0.2\ M_{\odot}$, but that the lack of observational evidence for
significant dust masses in Type Ia SNRs likely means that the
formation of carbon grains is suppressed by energetic photons and
electrons. They also find that the dust grains resulting from Type Ia
explosions have sizes below 0.01 microns and that they will likely not
survive the SN reverse shock. This implies that Type Ia SNe do not
inject a significant amount of dust into the ISM.

\section{Observational Evidence for Ejecta Dust in SNe}

\subsection{SN 1987A}
\label{1987A}

Observations of SN 1987A in the Large Magellanic Cloud (LMC) provided
the first direct evidence of dust condensation in SN ejecta and
allowed monitoring of the evolution of the SN's IR emission as it
transitioned to an SNR. The predictions that SN 1987A would form dust
a few hundred days post-explosion (Gehrz \& Nye 1987) were confirmed
when an IR excess was detected in the SN light curve (e.g. Moseley et
al. 1989). The fraction of the SN bolometric luminosity reradiated by
dust in the form of an IR continuum between 2-100 $\mu$m evolved from
2\% at day 260 post-explosion to as high as 83\% at day 775 (Wooden et
al. 1993). This coincided with the appearance of asymmetries in the
optical emission lines, providing firm evidence that dust condensed in
the ejecta of the SN explosion around day $\sim$ 500. The early IR
observations were consistent with $10^{-4} -
4\times10^{-3}\:M_{\odot}$ of dust, radioactively heated to a
temperature of $\sim$ 100 K, and consisting mostly of carbon grains,
as inferred from the absence of silicate features in the IR spectrum
(Wooden et al. 1993, Bouchet et al. 2004).

About ten years after explosion, as the blastwave of SN 1987A
encountered the equatorial ring, the IR emission became completely
dominated by pre-existing dust formed in the wind of the progenitor
star. Mid-IR monitoring, particularly with {\it Spitzer}, revealed
that the dust continuum emission could be explained by $\sim 10^{-5} -
10^{-6} \:M_{\odot}$ of mostly silicate grains, collisionally heated
by the X-ray emitting gas to a temperature of $\sim$ 180 K (e.g. Dwek
et al. 2010). During this time, the mid-IR emission from the
ejecta-condensed dust remained negligible.

\begin{figure}[t]
\sidecaption[t]
\includegraphics[scale=0.52]{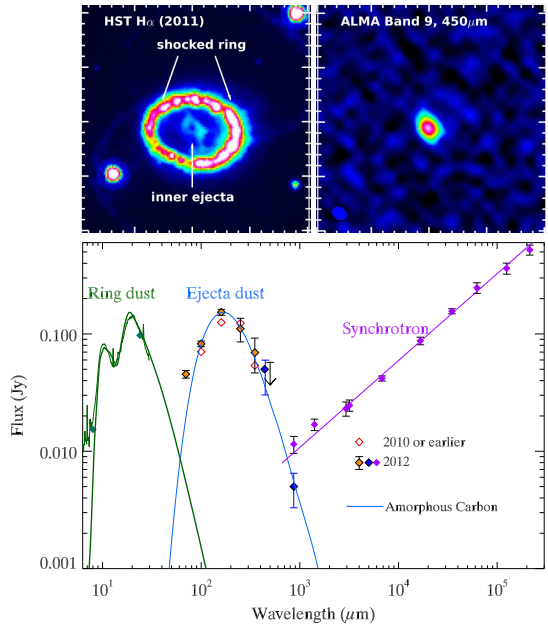}
\caption{{\it Top left}: {\it Hubble Space Telescope} H$\alpha$ line
  image of SN 1987A, showing the equatorial ring and the inner SN
  ejecta. {\it Top right}: ALMA 450 $\mu$m continuum image, confirming
  that the dust emission at far-IR wavelengths spatially coincides
  with the SN ejecta. {\it Bottom}: IR photometric and spectroscopic
  data for SN 1987A. The additional cold dust component emitting in
  the 100-350 $\mu$m range has been fit by 0.5 $M_{\odot}$ of dust at
  23 K (blue curve). Panels in the top row are from Indebetouw et
  al. (2014), and the bottom panel is from Matsuura et al. (2015).}
\label{1987afig}
\end{figure}

An additional cold dust component in SN 1987A was discovered by {\it
  Herschel} in mid- and far-IR (70-500 $\mu$m) observations. The
component was attributed to $\sim 0.5$ $M_{\odot}$ of SN-formed dust,
the largest mass ever detected in an SNR (Matsuura et al. 2015). The
IR spectral energy distribution, shown in Figure~\ref{1987afig}, was
fit by 0.5 $M_{\odot}$ of silicate and 0.3~$M_{\odot}$ of carbon
grains, emitting at a narrow temperature range of 20-25 K. Curiously,
these dust masses require a 100\% condensation efficiency for Si, and
exceed the expected nucleosynthetic yields for carbon. High spatial
resolution sub-millimeter observations with Atacama Large
Millimeter/submillimeter Array (ALMA) confirmed that the dust emission
originates from the central region of SN 1987A, coincident with the
location of the SN ejecta (Indebetouw et al. 2014). The dominant
heating source for the dust is likely the radioactive decay of
$^{44}$Ti.

The fact that the currently observed mass of ejecta dust is orders of
magnitude larger than the early estimates led several authors to
suggest that the increase in mass was caused by gradual coagulation
and accretion onto dust grains that formed at the early epochs, a few
hundred days after the explosion (Ercolano et al. 2007, Wesson et
al. 2015). Alternatively, Dwek et al. (2015) propose a scenario in
which $\sim$ 0.4~$M_{\odot}$ formed rapidly at the early epochs, but
was hidden in optically thick regions of the ejecta. The total dust
mass then evolves to 0.45~$M_{\odot}$ at late times, consisting of
$\sim$ 0.4~$M_{\odot}$ of silicate and 0.05 $M_{\odot}$ of carbon
dust. While the proposed scenario is not unique, it alleviates the
requirements for an unexpectedly large amount of Si, Mg, and C to be
locked up in dust, and explains the absence of silicate emission
features at the early epochs that would have been a result of
self-absorption in the optically thick clumps. In either scenario, the
large dust mass observed in SN 1987A is yet to encounter the SN
reverse shock that is expected to destroy a significant fraction of
the grains, so just how much of this dust will be injected into the
ISM remains an open question.

\subsection{Extragalactic Observations}

Extragalactic (which we define here as beyond the Milky Way and any
satellite galaxies, such as the MCs) observations in the IR are
limited to the SN phase. Nonetheless, there have been several
instances of dust being observed in extragalactic objects, both in
emission, where the continuum from hot dust causes an excess above the
normal SN light curve, and via indirect effects, such as asymmetric
line profiles from dust obscuration of receding material.

One of the early results in this field came from {\it Spitzer}
observations of SN 2003gd in NGC 628 (9.3 Mpc) by Sugerman et
al. (2006). The near-IR observations from 3.6 to 8 $\mu$m clearly show
a detection of the SN at day 499 that disappears by day 670. The
authors concluded that as much as 0.02 $M_{\odot}$ of dust had
condensed in the ejecta within $\sim 1$ yr post-explosion. However, a
later analysis by Meikle et al. (2007) came to the opposite
conclusion, namely that the near and mid-IR emission from SN 2003gd
could not be explained by newly formed dust, but instead was more
consistent with a light echo from pre-existing dust.

Kotak et al. (2009) obtained {\it Spitzer} photometric and
spectroscopic observations and optical spectra for several years
post-explosion in the Type IIP SN 2004et. They fit their data with a
three-component model which includes {\it both} emission from newly
formed dust in the ejecta and an IR light echo from dust in the host
galaxy. However, the amount of dust required to fit their observations
was only about 10$^{-4}$~$M_{\odot}$, several orders of magnitude
below the $\sim$ 0.1 $M_{\odot}$ needed, per core-collapse SN, to
reproduce the amount of dust in the universe. An extended analysis of
this object out to beyond day 2000 by Fabbri et al. (2011) confirmed
these results, without need for a large amount of dust formation.

Other observations of extragalactic SNe with {\it Spitzer} and other
instruments produce similar results. Nozawa et al. (2008) used {\it
  AKARI} observations of the Type Ib SN 2006jc to derive a mass of
only a few times 10$^{-3}$ $M_{\odot}$, despite theoretical
calculations showing that about 1 $M_{\odot}$ should form, based on
the temperature evolution of the SN. The Type IIP SN 2004dj, quite
close at only 3.5 Mpc, was observed several times with {\it Spitzer}
between days 98 and 1381 post-explosion, but Szalai et al. (2011) find
an upper limit of only 8 $\times 10^{-4}$ $M_{\odot}$. To date, no
significant amount of mass has ever been seen in an extragalactic SN,
though a caveat to this is that {\it Spitzer} may not be sensitive to
a cold dust component, which could peak at wavelengths $>40$ $\mu$m,
or that the dust emitting region may be optically thick (see
Section~\ref{1987A}). While both {\it Spitzer} and {\it Herschel} had
capabilities beyond this wavelength, their spatial resolution at long
wavelengths made observations of SNe in other galaxies
impossible. Once fully operational, ALMA may offer the possibility of
detecting cold dust from distant SNe in the years following explosion.

\section{Observational Evidence for Ejecta Dust in SNRs}

\subsection{Ejecta Dust in Shell-Type SNRs}

\begin{figure}[t]
%\sidecaption[t]
\includegraphics[scale=8.5]{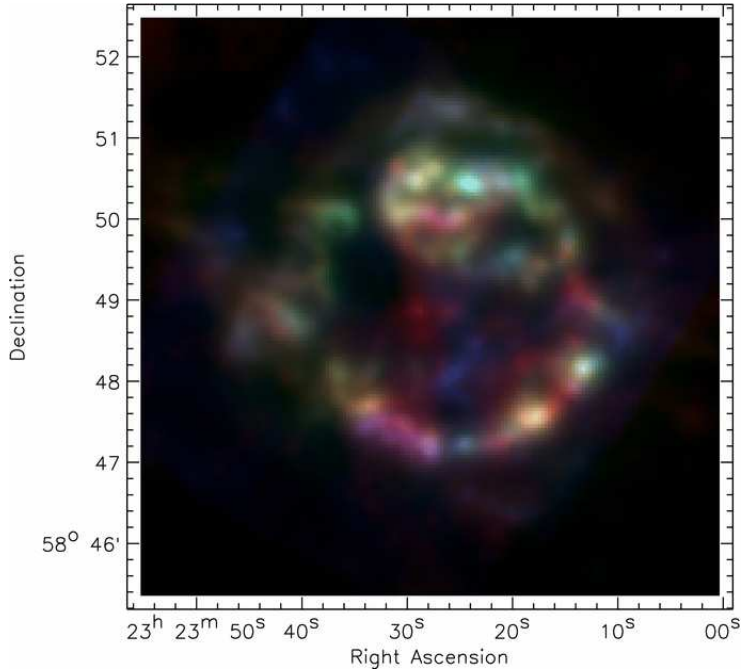}
\caption{A three-color image of Cas A (from Arendt et al. 2014),
  showing the dust continuum emission at 11.8 $\mu$m in blue, 20.8
  $\mu$m in green, and 70 $\mu$m in red. The various colors seen in
  the image reflect the spatial distribution of distinct species of
  dust that condensed in different layers of the SN ejecta.}
\label{casa}
\end{figure}

Most of the evidence for dust in the ejecta of SNRs has come from the
remnants of core-collapse SNe, as would be expected from theoretical
calculations (see Section~\ref{formation}). Perhaps the best example
is Cas A, where Barlow et al. (2010) use {\it Spitzer} and {\it
  Herschel} observations to separate out the warm and cold dust
components. The warm component arises from dust in the ISM heated by
the forward shock, but the cold component, with a measured mass of
0.075 $M_{\odot}$, is attributed to dust formed in the
ejecta. However, this cold dust is located interior to and has not yet
encountered the reverse shock, so the amount that will ultimately
survive is still unknown (see Section~\ref{destruction}). Arendt et
al. (2014) decomposed the IR emission into several different dust
components corresponding to various ejecta products in the remnant
(see Figure~\ref{casa}).

Other core-collapse remnants have yielded similar results. Sandstrom
et al. (2009) find only $\sim 10^{-3}$ $M_{\odot}$ in the Small
Magellanic Cloud remnant 1E0102-72.3. A survey of four LMC remnants by
Williams et al. (2006) found $< 0.1$ $M_{\odot}$ of dust in all of
those objects, and they attribute most of that to swept-up ISM
dust. They also point out that the morphology of the remnants in the
IR more closely matches the forward-shocked emission seen in X-rays,
rather than the SN ejecta.

Searches for dust in remnants of Type Ia SNe have turned up even
smaller amounts of dust. Borkowski et al. (2006) examined several
remnants in the LMC, and found no significant amounts of dust
associated with the ejecta. Gomez et al. (2012b) report {\it Herschel}
observations of Kepler's SNR and Tycho's SNR, and find no evidence for
a cool component of dust associated with the ejecta in either remnant.

\subsection{Ejecta Dust Around Pulsar Wind Nebulae}

Firm evidence for the existence of SN-formed dust is also found in
SNRs that contain pulsar-generated wind nebulae. In the early stages
of evolution of these systems, before the SN reverse shock has
propagated towards the center of the SNR, the expanding pulsar wind
nebula (PWN) drives a shock into the inner SN ejecta. Any dust that
might have condensed in the ejecta can then be heated to IR-emitting
temperatures, either collisionally by the shocked gas, or radiatively
by the high-energy synchrotron emission from the PWN. The heating
source provided by the central PWN therefore allows us to observe and
study pristine SN dust before it has been processed by the SN reverse
shock.

The most famous example of a system that exhibits this property is the
Crab Nebula, whose PWN sweeps up SN ejecta material observed in the
form of bright optical and IR filaments. The first observational
evidence for dust in the Crab Nebula was found in the form of an IR
excess above the synchrotron power-law in the integrated SED of the
PWN, and attributed to $10^{-3}-10^{-2}\:M_{\odot}$ of carbon and/or
silicate dust emitting at a temperature of $\sim$ 70 K. Further
evidence was found in the form of dust absorption features across the
Crab's optical ejecta filaments.

The high values of the dust-to-gas mass ratio and the spatial
correlation between the extinction features and low ionization
emission lines in the filaments suggested that the dust in the Crab
Nebula condensed from the SN ejecta, rather than originating from
swept-up circumstellar material. {\it Spitzer} mid-IR spectroscopy
confirmed that the dust continuum emission spatially correlates with
the brightest filaments, and showed that the spectrum is mostly
featureless (Temim et al. 2012b). The emission spectrum was fit by
$(1.2-12) \times 10^{-3}$ $M_{\odot}$ of warm dust, equally well
described by either silicate or carbonaceous grains, radiating at a
temperature of 55 $\pm$ 4 K and 60 $\pm$ 7 K, respectively.

Far-IR {\it Herschel} observations of the Crab Nebula revealed the
presence of a much more massive colder dust emission component ($\sim$
30 K) described by $0.24^{+0.32}_{-0.08}$ $M_{\odot}$ of silicate or
0.11$^{0.01}_{0.01}$ $M_{\odot}$ of carbon dust (Gomez et
al. 2012a). Temim \& Dwek (2013) modeled the mid and far-IR SED using
a physical dust heating model, in which the broadband emission from
the PWN radiatively heats dust grains with a continuous size
distribution. They found that the resulting continuous temperature
distribution of the grains leads to a reduction in the total dust mass
(0.019 - 0.13 $M_{\odot}$), more in line with the expected
nucleosynthetic yields for the Crab Nebula. On the other hand, Owen \&
Barlow (2015) used a photoionization and radiative transfer model to
show that the dust mass can be several times higher (up to 0.47
$M_{\odot}$) if the grains reside in dense clumps. Both models suggest
that the dust mass distribution is dominated by relatively large
grains ($>$ 0.1 $\mu$m), consistent with what is expected for a Type
IIP SN.

\begin{figure}
\includegraphics[scale=0.55]{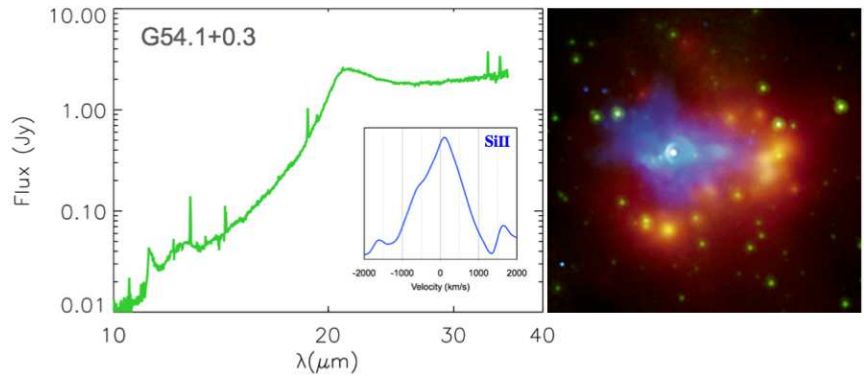}
\caption{{\it Left}: The \textit{Spitzer} mid-IR spectrum of the shell
  surrounding the pulsar wind nebula in SNR G54.1+0.3. A zoomed in
  velocity profile of the [Si II] 34.8 $\mu$m line is shown in blue in
  the inset. The dust spectrum shows mostly continuum emission, with a
  distinct broad feature peaking at $\sim$ 21~$\mu$m, arising from the
  same dust species as the ejecta dust in Cas A. The highly broadened
  emission line profiles, as shown by the Si line, indicate shell
  expansion velocities of $\sim$ 1000 km s$^{-1}$ and confirm that the
  emission originates from SN ejecta. {\it Right}: A three-color
  composite image of SNR G54.1+0.3, showing the \textit{Chandra} X-ray
  emission in blue, and \textit{Spitzer} 8 $\mu$m and 24 $\mu$m
  emission in green and red, respectively. The pulsar wind nebula
  emitting in the X-rays is surrounded by IR emission from SN ejecta
  and dust. The dust is being heated both by the pulsar wind, and the
  surrounding stars, as it blows past them. The IR point sources that
  stand out in yellow arise from the hotter dust in the immediate
  vicinity of the stars.  Image credit: X-ray: NASA/CXC/SAO/T.Temim et
  al. 2010; IR: NASA/JPL-Caltech}
\label{g54}
\end{figure}

Besides the Crab Nebula, three other PWNe show surrounding IR emission
that likely originates from SN-formed dust; B0540-69.3 in the LMC
(Williams et al. 2008), G54.1+0.3 (Temim et al. 2010), and Kes 75
(Temim et al. 2012a). {\it Spitzer} spectroscopy of B0540-69.3 shows a
dust continuum excess above the PWN synchrotron emission that can be
explained by a fairly small amount ($\sim$ 10$^{-3}$ $M_{\odot}$) of
dust, heated to a temperature of 50-65 K by the PWN shock (Williams et
al. 2008). A multi-wavelength study of G54.1+0.3 by Temim et
al. (2010) uncovered the unique nature of the IR shell surrounding the
PWN. The highly broadened (up to 1000 km~s$^{-1}$) emission lines
found in the {\it Spitzer} IRS spectrum (see Figure~\ref{g54}), which
spatially correlate with the shell emission, confirmed that the shell
is composed of SN ejecta and $\sim$ 0.1 $M_{\odot}$ of newly-formed
dust. Furthermore, the dust emission spectrum shows a distinct dust
feature at 21 $\mu$m, produced by the same dust species as the
SN-condensed dust in Cas A. Even more interestingly, the dust is being
heated not only by the PWN shock, but primarily by the O and B stars
that are members of the cluster in which the SN progenitor
existed. While the dust masses in these systems are relatively low,
especially considering that the grains have not yet been encountered
by the SN reverse shock that is expected to destroy a significant
fraction, they were derived from {\it Spitzer} mid-IR observations
that only sample the warmer grains. Far-IR emission has now been
detected from all three systems with {\it Herschel}, suggesting that
more dust resides at colder temperatures. Future studies and searches
for IR emission around other young PWNe will be important for
increasing the number of systems for which we can characterize the
properties and masses of unprocessed SN dust.

\section{Dust Destruction in Shocks}
\label{destruction}

\subsection{Destruction Mechanisms}

\begin{figure}[t]
\sidecaption[t]
\includegraphics[scale=0.4]{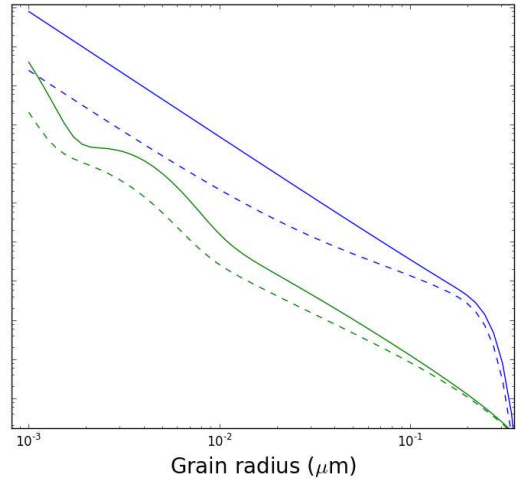}
\caption{The pre-shock (solid lines) and post-shock (dashed lines)
  dust grain size distributions of silicate and graphite dust grains
  1000 years post-shock. The density in this simulation was n$_{0} =
  2$ cm$^{-3}$, with an ion temperature of 5 keV. Silicate grains are
  shown in blue, while graphite is shown in green. Small grains are
  the most affected; see text for details. The vertical scales between
  silicate and graphite, reflecting the number of grains, are
  arbitrarily offset for ease of comparison, but the scales within
  each grain type are accurate.}
\label{sputtering}
\end{figure}

Dust grains are generally destroyed via collisions, either with gas
particles (sputtering) or with other grains (shattering). In
sputtering, grains are bombarded by energetic protons and other
ions. These collisions knock off a few atoms at a time, depending on
the energy of the ion, gradually eroding the grain surface. To first
order, the sputtering rate (i.e., the number of atoms knocked off per
collision) is independent of the size of the grain, meaning that
smaller grains are destroyed more quickly. This results in a
modification of the shape of the grain size distribution with time. In
Figure~\ref{sputtering}, we show an example of this, where the
pre-shock and post-shock grain size distributions for both silicate
and carbonaceous grains are shown after a processing time (years after
the shock passage) of 1000 years.

Thermal sputtering, described above, dominates the destruction of
grains for fast, non-radiative shocks (Draine \& Salpeter 1979). For
slower shocks, non-thermal sputtering, sometimes referred to as
``ballistic'' sputtering, is important as well. This sputtering arises
from the relative gas-grain motions in the post-shock region. Dust
grains generally have a low charge-to-mass ratio, and are largely
unaffected by the passage of the shock (in SNRs, the shock wave is
collisionless). The plasma around them, however, suddenly acquires a
velocity of (3/4)$v_{s}$, where $v_{s}$ is the shock speed. They are
eventually brought to rest in the frame of the gas due to the drag
force from collisions with ions, but this takes a non-negligible
amount of time, during which their sputtering rate is increased.

Finally, shattering of grains via grain-grain collisions reprocesses
grains by reducing their size. While not a destruction mechanism, per
se, since the mass remains in the solid dust phase, this process also
affects the grain size distribution, tilting it towards smaller grains
that are then more easily destroyed by sputtering. Jones et al. (1996) show
that for slow shocks of $< 200$ km s$^{-1}$, 5-15\% of the initial
mass in grains may be reprocessed via shattering to very small grain
sizes of less than 1 nm, and thus may be a significant source of
polycyclic aromatic hydrocarbons (PAHs) and other small grains.

\subsection{Destruction of SN Dust by the Reverse Shock}
\label{reverseshock}

While core-collapse SNe are capable of producing a significant amount
of dust from their ejecta, the ultimate survival of this dust
following the passage of the SN reverse shock determines whether they
are indeed major sources of dust in galaxies. The survival of dust
grains depends not only on the type of explosion (see Section 3.3),
but also on the clumpiness of SN ejecta, and the density structure of
the surrounding ISM. The theoretical models generally show that
40-100\% of the dust mass is destroyed in the encounter with the
reverse shock, and that grains with sizes below 0.1 $\mu$m are
completely sputtered away (e.g. Kozasa et al. 2009, Silvia et
al. 2010).

While the dust produced in Type Ia and IIb SNe is shown to be almost
completely destroyed, even for the highest gas densities and clumpy
ejecta, as much as 0.2 $\rm M_{\odot}$ may survive in Type IIP SNRs
due to the larger size of grains that condense in the ejecta (Kozasa
et al. 2009, Nozawa et al. 2011, Biscaro \& Cherchneff 2014).  We note
that for all SNRs in which significant dust masses have been
discovered to date (SN 1987A, Cas A, Crab Nebula, and G54.1$+$0.3),
the dust resides on the inside of the reverse shock and is yet to be
processed. So far, the only observational evidence for SN dust
surviving the reverse shock is the discovery of 0.02 $\rm M_{\odot}$
of dust in the interior of a 10,000-year-old SNR Sgr A East (Lau et
al. 2015).

\subsection{Observational Evidence for Destruction in SNRs}

Modeling the IR emission from individual remnants shows direct
evidence for the destruction of ISM dust via sputtering. Borkowski et
al. (2006) and Williams et al. (2006) used early {\it Spitzer} imaging
observations of young SNRs in the LMC to show that the observed fluxes
could not be modeled by standard, unmodified grain size
distributions. The fits required dust destruction of order $\sim
25-50$\% of the mass in grains encountered by the forward shock. Blair
et al. (2007) did the same for Kepler's SNR, where the high densities
encountered by the forward shock lead to $\sim$ 75\% of the grains
being destroyed. Arendt et al. (2010) observed Puppis A, an older
galactic remnant interacting with a molecular cloud, to reach a
similar conclusion. Since Puppis A is a large remnant, they were able
to place the {\it Spitzer} spectroscopic slits at various points
behind the shock and follow the evolution of the dust grain size
distribution. They found that the {\it Spitzer} spectra could only be
fit by removing 25\% of the mass in grains. A similar conclusion was
reached by Sankrit et al. (2014) in observations of the Cygnus Loop,
another older remnant occupying a large volume of the ISM. It is
important to note that grain destruction does not merely lower the IR
flux from the remnant, but also changes the shape of the spectrum, due
to small grains being preferentially destroyed.

{\it Spitzer} is now in its warm phase and no longer has spectroscopic
capabilities. However, the {\it James Webb Space Telescope} (JWST) is
scheduled for launch in 2018, and will have both imaging and
spectroscopic capabilities from $\sim 1-25~\mu$m. JWST's angular
resolution will be nearly an order of magnitude better than {\it
  Spitzer's}, and will allow studies like the one done on Puppis A to
be done on several remnants, both in the Galaxy and the MCs. The much
greater sensitivity means that fainter remnants can be studied as
well.

\subsection{Supernovae as Net Destroyers of Dust?}

Besides potentially being important sources of dust in galaxies, SNe
are also responsible for the majority of dust destruction in the
ISM. The efficiency of the destruction depends on the ISM conditions,
such as the gas density, homogeneity, metallicity, magnetic field, and
dust-to-gas mass ratio, as well as on the properties of the ISM dust
grains themselves.  Various theoretical models have been developed to
study the effect of SN blastwaves on dust in the ISM, and they
generally agree that SN dust destruction rates exceed the SN injection
rates by at least an order of magnitude (Jones et al. 1996). Slavin et
al. (2015) modeled dust destruction by radiative SNRs expanding in a
homogeneous, magnetized, warm phase of the ISM, and found that a
typical SNR in the Galaxy would be expected to destroy $\sim 3.7 \:
\rm M_{\odot}$ of silicate and $\sim 1.4 \: \rm M_{\odot}$ of carbon
dust, significantly more than a typical SN is expected to produce (see
Section~\ref{formation}). This discrepancy implies that dust in the
ISM is either shielded somehow, or that a large fraction grows in
dense molecular clouds.

In a recent study, Temim et al. (2015) estimated the global dust
destruction rate by a complete sample of SNRs in the MCs by combining
the observational constraints on the dust-to-gas mass ratio and gas
density around each SNR, and the theoretically determined dust
destruction efficiencies. They calculated the total amount of dust
each SNR will destroy throughout its evolution, and found that on
average, 7.7$\pm$2.7 $\rm M_{\odot}$ is destroyed by an SNR in the LMC
and 3.3$\pm$1.8 $\rm M_{\odot}$ in the SMC. Lakicevic et al. (2015)
estimated the amount of dust destruction in the LMC by comparing the
dust mass surface densities inside and outside the SNRs with their
surrounding media. They found evidence for dust destruction in six
SNRs, with an average of 6.5 $\rm M_{\odot}$ destroyed. Ochsendorf et
al. (2015) find that dust processing by SNRs inside superbubbles in
the LMC increases the dust destruction efficiency even further, since
dust in continually replenished through cloud evaporations inside the
bubbles and destroyed by subsequent SNe. These studies support the
idea that SNe are net destroyers of dust in the ISM, at least in
present-day galaxies.

\section{Conclusions}

In light of recent IR observations of young SNRs, particularly SN
1987A, there is no doubt that SNe can form significant quantities of
dust. However, the properties and amount of dust produced in different
progenitor types, as well as the fraction of dust that ultimately
survives the remnant phase to be injected into the ISM, are not well
known. Additionally, since the SNR shock waves destroy significant
quantities of ISM dust, it is likely that SNe, as a whole, are net
destroyers of dust in galaxies. Significant progress in this field
will be made upon the successful launch of JWST, a mid-IR observatory
with a resolution about an order of magnitude better than Spitzer and
sensitivity several hundred times higher. JWST will allow the followup
of extragalactic SNe for tens of years in some cases, and will serve
as a valuable monitor on SN 1987A’s continuing evolution. The superior
resolution will also further disentangle the various dust components
in SNRs in both the Galaxy and the MCs, and allow studies of the
efficiency of dust destruction by SNR shocks. A combination of
observations of both SNe and SNRs, along with theoretical modeling,
will be necessary to solve the mystery of the origin of dust in the
universe.

%%%%%%%%%%%%%%%%%%%%%%%% referenc.tex %%%%%%%%%%%%%%%%%%%%%%%%%%%%%%
% sample references
% %
% Use this file as a template for your own input.
%
%%%%%%%%%%%%%%%%%%%%%%%% Springer-Verlag %%%%%%%%%%%%%%%%%%%%%%%%%%
%
% BibTeX users please use
% \bibliographystyle{}
% \bibliography{}

\begin{thebibliography}{99.}%

\bibitem{phys-journal} R.G. Arendt, E. Dwek, W.P. Blair, P. Ghavamian,
  U. Hwang, K.S. Long, et al. 2010, Spitzer Observations of Dust
  Destruction in the Puppis A Supernova Remnant, ApJ, {\bf 725, 585}

\bibitem{phys-journal} R.G. Arendt, E. Dwek, G. Kober, J. Rho, \&
  U. Hwang 2014, Interstellar and Ejecta Dust in the Cas A Supernova
  Remnant, ApJ, {\bf 786, 55}

\bibitem{phys-journal} M.J. Barlow, O. Krause, B.M. Swinyard,
  B. Sibthorpe, M.-A. Besel, R. Wesson, et al. 2010, A Herschel PACS
  and SPIRE study of the dust content of the Cassiopeia A supernova
  remnant, A\&A, {\bf 518, 138}

\bibitem{phys-journal} R. Becker \& W. Doring 1935, K{\'i}netische
  Behandlung der Keimbildung in {\"u}bers{\"a}ttigten D{\"a}mpfen,
  Annalen der Physik, {\bf 24, 719}

%\bibitem{phys-journal} S. Bianchi, \& R. Schneider, 2007, MNRAS,
%  \textbf{378, 973}

\bibitem{phys-journal} C. Biscaro, \& I. Cherchneff, 2014, Molecules
  and dust in Cassiopeia A. I. Synthesis in the supernova phase and
  processing by the reverse shock in the clumpy remnant, A\&A,
  \textbf{564, A25}

\bibitem{phys-journal} W.P. Blair, P. Ghavamian, K.S. Long,
  B.J. Williams, K.J. Borkowski, S.P. Reynolds, et al. 2007, Spitzer
  Space Telescope Observations of Kepler's Supernova Remnant: A
  Detailed Look at the Circumstellar Dust Component, ApJ, {\bf 662,
    998}

%\bibitem{phys-journal} M. Bocchio, A.~P. Jones, \& J.~D. Slavin, 2014,
%  A\&A, \textbf{570, A32}

\bibitem{phys-journal} K.J. Borkowski, B.J. Williams, S.P. Reynolds,
  W.P. Blair, P. Ghavamian, R. Sankrit, et al. 2006, Dust Destruction
  in Type Ia Supernova Remnants in the Large Magellanic Cloud, ApJ,
  {\bf 642, 141}

\bibitem{phys-journal} P. Bouchet, J.M. De Buizer, N.B. Suntzeff,
  J.I. Danziger, T.L.Hayward, C.M. Telesco, et al. 2004,
  High-Resolution Mid-infrared Imaging of SN 1987A, ApJ, \textbf{611,
    394}

%\bibitem{phys-journal} I. Cherchneff, \& E. Dwek, 2010, ApJ,
%  \textbf{713, 1}

\bibitem{phys-journal} B.T. Draine \& Salpeter, E.E., 1979,
  Destruction mechanisms for interstellar dust, ApJ, {\bf 231, 77}

\bibitem{phys-journal} B.~T. Draine, 2009, Cosmic Dust - Near and Far
  ASP Conference Series, Vol. 414, proceedings of a conference held
  8-12 September 2008 in Heidelberg, Germany. Edited by Thomas
  Henning, Eberhard Grün, and Jürgen Steinacker. San Francisco:
  Astromomical Society of the Pacific, \textbf{414, 453}

\bibitem{phys-journal} E. Dwek, F. Galliano, \& A. Jones, 2009, Cosmic
  Dust - Near and Far ASP Conference Series, Vol. 414, proceedings of a conference held 8-12 September 2008 in Heidelberg, Germany. Edited by Thomas Henning, Eberhard Grün, and Jürgen Steinacker. San Francisco: Astromomical Society of the Pacific, \textbf{414, 183}

\bibitem{phys-journal} E. Dwek, R.G. Arendt, P. Bouchet, D.N. Burrows,
  P. Challis, J.I. Danziger, et al. 2010, Five Years of Mid-infrared
  Evolution of the Remnant of SN 1987A: The Encounter Between the
  Blast Wave and the Dusty Equatorial Ring, ApJ, \textbf{722, 425}

\bibitem{phys-journal} M. Elvis, M. Marengo, \& M. Karovska, 2002,
  Smoking Quasars: A New Source for Cosmic Dust, ApJL, \textbf{567,
    L107}

\bibitem{phys-journal} B. Ercolano, M.~J. Barlow, \&
  B.~E.~K. Sugerman, 2007, Dust yields in clumpy supernova shells: SN
  1987A revisited, MNRAS, \textbf{375, 753}

\bibitem{phys-journal} J. Fabbri, M. Otsuka, M.J. Barlow,
  J.S. Gallagher, R. Wesson, B.E.K. Sugerman, et al. 2011, The effects
  of dust on the optical and infrared evolution of SN 2004et, MNRAS,
  {\bf 418, 1285}

%\bibitem{phys-journal} O.D. Fox, et al., 2010, ApJ, {\bf 725, 1768}

\bibitem{phys-journal} C. Gall, J. Hjorth, \& A.~C. Andersen, 2011,
  Production of dust by massive stars at high redshift, A\&ARv,
  \textbf{19, 43}

\bibitem{phys-journal} R.~D. Gehrz,\& E.~P. Ney, 1987, On the
  Possibility of Dust Condensation in the Ejecta of Supernova 1987a,
  Proceedings of the National Academy of Science, \textbf{84, 6961}

\bibitem{phys-journal} H.~L. Gomez, O. Krause, M.J. Barlow,
  B.M. Swinyard, P.J. Owen, C.J.R. Clark, et al. 2012a, A Cool Dust
  Factory in the Crab Nebula: A Herschel Study of the Filaments, ApJ,
  \textbf{760, 96}

\bibitem{phys-journal} H.L. Gomez, C.J.R. Clark, T. Nozawa, O. Krause,
  E.L. Gomez, M. Matsuura, et al. 2012b, Dust in historical Galactic
  Type Ia supernova remnants with Herschel, MNRAS, {\bf 420, 3557}

%\bibitem{phys-journal} J.~J. Hester, 2008, ARAA, {\bf 46, 127}

%\bibitem{phys-journal} H. Hirashita, T. Nozawa, T.~T. Takeuchi, \&
%  T. Kozasa, 2008, MNRAS, \textbf{384, 1725}

\bibitem{phys-journal} R. Indebetouw, M. Matsuura, E. Dwek,
  G. Zanardo, M.J. Barlow, M. Baes, et al. 2014, Dust Production and
  Particle Acceleration in Supernova 1987A Revealed with ALMA, ApJL,
  \textbf{782, L2}

\bibitem{phys-journal} A.~P. Jones, A.~G.~G.~M. Tielens, \&
  D.~J. Hollenbach, 1996, Grain Shattering in Shocks: The Interstellar
  Grain Size Distribution, ApJ, \textbf{469, 740}

\bibitem{phys-journal} R. Kotak, W.P.S. Meikle, D. Farrah,
  C.L. Gerardy, R.J. Foley, S.D. Van Dyk, et al. 2009, Dust and The
  Type II-Plateau Supernova 2004et, ApJ, {\bf 704, 306}

\bibitem{phys-journal} T. Kozasa, T. Nozawa, N. Tominaga, H. Umeda,
  K. Maeda, K. Nomoto 2009, Cosmic Dust - Near and Far ASP Conference
  Series, Vol. 414, proceedings of a conference held 8-12 September
  2008 in Heidelberg, Germany. Edited by Thomas Henning, Eberhard
  Grün, and Jürgen Steinacker. San Francisco: Astromomical Society of
  the Pacific, \textbf{414, 43}

\bibitem{phys-journal} R.~M. Lau, T.~L. Herter, M.~R. Morris, Z. Li,
  \& J.~D. Adams, 2015, Old supernova dust factory revealed at the
  Galactic center, Science, \textbf{348, 413}

%\bibitem{phys-journal} A. Loll, 2010, Ph.D.~Thesis, Arizona State
%  University

%\bibitem{phys-journal} L.~B. Lucy, I.~J. Danziger, C. Gouiffes, \&
%  P. Bouchet, 1989, IAU Colloq.~120: Structure and Dynamics of the
%  Interstellar Medium, \textbf{350, 164}

%\bibitem{phys-journal} J.S. Mathis, W. Rumpl, \& K.H. Nordsieck, 1977,
%  ApJ, {\bf 217, 425}

\bibitem{phys-journal} M. Matsuura, E. Dwek, M.J. Barlow, B. Babler,
  M. Baes, M. Meixner, et al. 2015, A Stubbornly Large Mass of Cold
  Dust in the Ejecta of Supernova 1987A, ApJ, \textbf{800, 50}

\bibitem{phys-journal} W.P.S. Meikle, S. Mattila, A. Pastorello,
  C.L. Gerardy, R. Kotak, J. Sollerman, et al. 2007, A Spitzer Space
  Telescope Study of SN 2003gd: Still No Direct Evidence that
  Core-Collapse Supernovae are Major Dust Factories, ApJ, {\bf 665,
    608}

\bibitem{phys-journal} T. Nozawa, T. Kozasa, N. Tominaga, I. Sakon,
  M. Tanaka, T. Suzuki, et al. 2008, Early Formation of Dust in the
  Ejecta of Type Ib SN 2006jc and Temperature and Mass of the Dust,
  ApJ, {\bf 684, 1343}

\bibitem{phys-journal} T. Nozawa, K. Maeda, T. Kozasa, M. Tanaka,
  K. Nomoto, \& H. Umeda, 2011, Formation of Dust in the Ejecta of
  Type Ia Supernovae, ApJ, \textbf{736, 45}

\bibitem{phys-journal} B.~B. Ochsendorf, A.~G.~A. Brown, J. Bally, \&
  A.~G.~G.~M. Tielens, 2015, Nested Shells Reveal the Rejuvenation of
  the Orion-Eridanus Superbubble, ApJ, \textbf{808, 111}

\bibitem{phys-journal} P.~J. Owen, \& M.~J. Barlow 2015, The Dust and
  Gas Content of the Crab Nebula, ApJ, \textbf{801, 141}

\bibitem{phys-journal} W.T. Reach, J. Rho, A. Tappe, T.G. Pannuti,
  C.L. Brogan, E.B. Churchwell, et al. 2006, A Spitzer Space Telescope
  Infrared Survey of Supernova Remnants in the Inner Galaxy, AJ, {\bf
    131, 1479}

%\bibitem{phys-journal} S.P. Reynolds, et al. 2007, ApJ, {\bf 668, 135}

%\bibitem{phys-journal} J. Rho, et al. 2003, ApJ, {\bf 592, 299}

%\bibitem{phys-journal} J. Rho, et al. 2009, ApJ, {\bf 700, 579}

\bibitem{phys-journal} K.M. Sandstrom, A.D. Bolatto, D. Alberto,
  S. Stanimirovi{\'c}, J. van Loon, J.D.T. Smith, 2009, Measuring Dust
  Production in the Small Magellanic Cloud Core-Collapse Supernova
  Remnant 1E 0102.2-7219, ApJ, {\bf 696, 2138}

\bibitem{phys-journal} R. Sankrit, J.C. Raymond, M. Bautista,
  T.J. Gaetz, B.J.  Williams, W.P. Blair, et al. 2014, Spitzer IRS
  Observations of the XA Region in the Cygnus Loop Supernova Remnant,
  ApJ, {\bf 787, 3}

\bibitem{phys-journal} A. Sarangi, \& I. Cherchneff, 2015,
  Condensation of dust in the ejecta of Type II-P supernovae, A\&A,
  \textbf{575, A95}

\bibitem{phys-journal} D.~W. Silvia, B.~D. Smith, \& J.~M. Shull,
  2010, Numerical Simulations of Supernova Dust
  Destruction. I. Cloud-crushing and Post-processed Grain Sputtering,
  ApJ, \textbf{715, 1575}

\bibitem{phys-journal} J.~D. Slavin, E. Dwek, \& A.~P. Jones, 2015,
  Destruction of Interstellar Dust in Evolving Supernova Remnant Shock
  Waves, ApJ, \textbf{803, 7}

\bibitem{phys-journal} N. Smith, R.J. Foley, A.V. Filippenko, 2008,
  Dust Formation and He II λ4686 Emission in the Dense Shell of the
  Peculiar Type Ib Supernova 2006jc, ApJ, {\bf 680, 568}

\bibitem{phys-journal} B.E. Sugerman, B. Ercolano, M.J. Barlow,
  A.G.G.M. Tielens, G.C. Clayton, A.A. Zijlstra, et al. 2006,
  Massive-Star Supernovae as Major Dust Factories, Science, {\bf 313,
    196}

\bibitem{phys-journal} T. Szalai, J. Vink{\'o}, Z. Balog,
  A. G{\'a}sp{\'a}r, M. Block, L.L. Kiss, 2011, Dust formation in the
  ejecta of the type II-P supernova 2004dj, A\&A, {\bf 527, 61}

\bibitem{phys-journal} T. Temim, P. Slane, S.~P. Reynolds,
  J.~C. Raymond, \& K.~J. Borkowski, 2010, Deep Chandra Observations
  of the Crab-like Pulsar Wind Nebula G54.1+0.3 and Spitzer
  Spectroscopy of the Associated Infrared Shell, ApJ, \textbf{710,
    309}


\bibitem{phys-journal} T. Temim, P. Slane, R.~G. Arendt, \& E. Dwek,
  2012a, Infrared and X-Ray Spectroscopy of the Kes 75 Supernova Remnant Shell: Characterizing the Dust and Gas Properties, ApJ, \textbf{745, 46}


\bibitem{phys-journal} T. Temim, G. Sonneborn, E. Dwek, R.G. Arendt,
  R.D. Gehrz, P. Slane, et al. 2012b, Properties and Spatial
  Distribution of Dust Emission in the Crab Nebula, ApJ, \textbf{753,
    72}


\bibitem{phys-journal} T. Temim, \& E. Dwek, 2013, The Importance of
  Physical Models for Deriving Dust Masses and Grain Size
  Distributions in Supernova Ejecta. I. Radiatively Heated Dust in the
  Crab Nebula, ApJ, \textbf{774, 8}

\bibitem{phys-journal} T. Temim, E. Dwek, K. Tchernyshyov, M.L. Boyer,
  M. Meixner, C. Gall, et al. 2015, Dust Destruction Rates and
  Lifetimes in the Magellanic Clouds, ApJ, \textbf{799, 158}

%\bibitem{phys-journal} P. Todini,\& A. Ferrara, 2001, MNRAS,
%  \textbf{325, 726}

\bibitem{phys-journal} R. Wesson, R., M.~J. Barlow, M. Matsuura, \&
  B. Ercolano, 2015, The timing and location of dust formation in the
  remnant of SN 1987A, MNRAS, \textbf{446, 2089}

\bibitem{phys-journal} B.J. Williams, K.J. Borkowski, S.P. Reynolds,
  W.P. Blair, P. Ghavamian, S.P. Hendrick, et al., 2006, Dust
  Destruction in Fast Shocks of Core-Collapse Supernova Remnants in
  the Large Magellanic Cloud, ApJ, {\bf 652, 33}

\bibitem{phys-journal} B.J. Williams, K.J. Borkowski, S.P. Reynolds,
  J.C. Raymond, K.S. Long, J. Morse, et al., 2008, Ejecta, Dust, and
  Synchrotron Radiation in SNR B0540-69.3: A More Crab-Like Remnant
  than the Crab, ApJ, \textbf{687, 1054}

\bibitem{phys-journal} B.J. Williams, K.J. Borkowski, S.P. Reynolds,
  P. Ghavamian, J.C. Raymond, K.S. Long, et al., 2011, Dusty Blast
  Waves of Two Young Large Magellanic Cloud Supernova Remnants:
  Constraints on Post-shock Compression, ApJ, {\bf 729, 65}

\bibitem{phys-journal} B.J. Williams, K.J. Borkowski, S.P. Reynolds,
  P. Ghavamian, W.P. Blair, K.S. Long, et al. 2012, Dust in a Type Ia
  Supernova Progenitor: Spitzer Spectroscopy of Kepler's Supernova
  Remnant, ApJ, {\bf 755, 3}

\bibitem{phys-journal} D.~H. Wooden, D.M. Rank, J.D. Bregman,
  F.C. Witteborn, A.G.G.M. Tielens, M. Cohen, et al. 1993, Airborne
  spectrophotometry of SN 1987A from 1.7 to 12.6 microns - Time
  history of the dust continuum and line emission, ApJS, \textbf{88,
    477}

%\bibitem{phys-journal} V. Zubko, E. Dwek, \& R.G. Arendt 2004, ApJS,
  {\bf 152, 211}

\end{thebibliography}
%

\end{document}